\documentclass[reprint,twocolumn,showpacs,superscriptaddress,10pt,nofootinbib, amsmath,amssymb, aps, prd]{revtex4-1}
\usepackage{graphicx}
\usepackage{xspace}
\usepackage[english]{babel}
\usepackage[usenames,dvipsnames]{color}     
\usepackage{comment}
\usepackage[font=small]{caption}
\usepackage{subcaption}
\usepackage{algorithm}
\usepackage{algpseudocode} 
\usepackage[backref=page]{hyperref}

\definecolor{linkcolor}{rgb}{.8,0,0}
\definecolor{urlcolor}{rgb}{0,0,.7}
\definecolor{citecolor}{rgb}{0,.5,0}
\definecolor{acrocolor}{rgb}{0,0,.7}
\hypersetup{bookmarksopen,colorlinks=true}
\hypersetup{pdfstartview=FitH}
\hypersetup{linktocpage=true,bookmarksnumbered=true}
\hypersetup{plainpages=false,breaklinks=true}
\hypersetup{linkcolor=linkcolor,citecolor=citecolor,urlcolor=urlcolor}

\newcommand{\Finesse}{\textsc{Finesse}\xspace}

\newcommand{\I}{\ensuremath{\mathrm{i}}\xspace}
\newcommand{\maxtem}{\ensuremath{\mathcal{O}_{max}}\xspace}

\definecolor{cadmiumgreen}{rgb}{0.0, 0.8, 0.24}

\newcommand{\be}{\begin{equation}}
\newcommand{\ee}{\end{equation}}

\begin{document}

\title{Fast Simulation of Gaussian-Mode Scattering for Precision Interferometry}
	
\author{D. Brown}
\email{Corresponding author: ddb@star.sr.bham.ac.uk}
\address{University of Birmingham, Edgbaston, B15 2TT, UK}
\author{R. J. E. Smith}
\address{LIGO, California Institute of Technology, Pasadena, CA 91125, USA}
\author{A. Freise}
\address{University of Birmingham, Edgbaston, B15 2TT, UK}

\begin{abstract}
Understanding how laser light scatters from realistic mirror surfaces
is crucial for the design, commissioning and operation of precision
interferometers, such as the current and next generation of
gravitational-wave detectors. Numerical simulations are indispensable
tools for this task but their utility can in practice be limited by
the computational cost of describing the scattering process. In this
paper we present an efficient method to significantly reduce the
computational cost of optical simulations that incorporate
scattering. This is accomplished by constructing a near optimal
representation of the complex, multi-parameter 2D overlap integrals that
describe the scattering process (referred to as a \textit{reduced order
quadrature}). We demonstrate our technique by simulating a
near-unstable Fabry-Perot cavity and its control signals using similar
optics to those installed in one of the LIGO gravitational-wave
detectors. We show that using reduced order quadrature reduces the
computational time of the numerical simulation from days to minutes
(a speed-up of $\approx2750\times$) whilst incurring negligible
errors. This significantly increases the feasibility of modelling
interferometers with realistic imperfections to overcome current
limits in state-of-the-art optical systems. Whilst we focus on the Hermite-Gaussian
basis for describing the scattering of the optical fields, our method
is generic and could be applied with any suitable basis. An implementation 
of this reduced order quadrature method is provided in the open source
interferometer simulation software \Finesse.
\end{abstract}

\maketitle 

\section{Introduction}

Laser interferometers have long been an exceptional tool for enabling high precision measurements.
With ever increasing demands on their performance, new techniques and tools have been developed to
design and build the next-generation of instruments. This has especially been true in the development
of gravitational-wave detectors over the last several decades
~\cite{Grote08, Sigg08, acernese08, Takahashi04b}. 
Such ground-based gravitational-wave
detectors are based on a Michelson 
interferometer and are enhanced with Fabry-Perot cavities.
Detecting gravitational waves is still one of the major challenges in
experimental physics, and the interferometers used include numerous new optical technologies 
to reach unprecedented displacement sensitivities beyond
$10^{-19}\,\rm{m}/\sqrt{\rm{Hz}}$.

Some of these detectors are currently being upgraded to 
have a ten-fold increase in sensitivity using a much higher
circulating power \cite{AdvancedLIGO15, AdvancedVirgo15}.
To achieve their target performance the detectors undergo several
years of commissioning, during which the interferometers are
carefully tested and improved towards their designed operational state.
Numerical simulations are important tools 
to diagnose causes of any unexpected behaviour
seen during commissioning; to suggest solutions to potential problems
and for advising the design of detector upgrades.
Hence, there is a long history of developing and using dedicated 
optical simulation tools for the commissioning and design of
gravitational-wave detectors~\cite{STAIC, GWIC, Siesta-paper,
  e2e_2000}.


One of the key aspects for the current instruments is the 
high circulating laser power, up to hundreds of kiloWatts, required
for a broadband reduction of shot-noise. It has been recognised
for some time that the thermal deformations of the optics due to spurious
absorption can degrade the performance of the interferometers~\cite{vinet09}. Numerical models 
have been used extensively in the investigation of
such problems and in the development of mitigating
solutions (for example~\cite{Vinet1992, Beausoleil03}). Thermally
induced distortions and other effects related to the laser beam shape
are still limiting factors of the instruments today and are concerns for
the design of future detectors~\cite{ET-0106C-10}.
Furthermore, similar effects can limit the performance of other optical precision
measurements such as optical clocks~\cite{Schibli08} or the optical readout of atomic
systems~\cite{Dickerson13}. Mitigation strategies for beam shape distortions in complex
interferometers are actively being developed and require accurate
numerical models for their design and development.

Initially, the simulation tools for investigating distorted beams used a grid-based
field description. Beam distortions can also be modeled effectively
using an expansion into spatial cavity eigenmodes~\cite{freise10}, such as
Hermite-Gauss modes.
The interaction of the beam shape with a distorted optical surface 
often requires the computation of a scattering matrix based on measured
or simulated profiles of the distorted surface. This is always true
for mode-based simulations programs but is also required for
grid-based codes when specific shapes of the beam are important, for
example, for the investigations of parametric
instabilities~\cite{BSV01,Evans2010665}.
If this matrix has to be re-generated, for example when the effects
of a change of a surface shape is being investigated,
this element of the computation can dominate the total time required
for the entire simulation. A prominent example is that when the circulating laser power
within the LIGO interferometers thermally warps the mirror surfaces 
changing the shape of the laser beams and requiring a re-calculation
of many scattering matrices. Including this effect can increase
the computation time from minutes to days.


Some of us are providing  numerical simulation support for 
the commissioning of the LIGO interferometers~\cite{T1300954}.
We use our own simulation tool \Finesse~\cite{Freise04} and are maintaining parameter files for the
detectors~\cite{aligo_file}. Commissioning tackles the unexpected
behaviour of the interferometers and must take into account the sometimes rapid
progress of the experimental setup. Therefore support provided with
numerical models must fulfil two criteria: a) we must be able to
accurately model the current experimental setup in the presence of
distortions and deviations from the design and b) we must be
able to provide a quick response to new questions to inform the
management of the activities on site in real time. \Finesse is a
frequency-domain tool, using Hermite-Gauss modes to describe beam
distortions and is thus ideally suited as a rapid and accurate tool.

Our investigations with numerical tools typically consist of a
sequence of different subtasks, sometimes using different tools,
alternating with an expert review of intermediate or preliminary
results. This is a very different pattern of tasks to those that benefit
from a computer cluster or super computer. Instead, our work
requires lightweight and flexible tools with computing times up to
minutes or hours. Because of this, strategies to ameliorate the run time of simulations
are of high importance to provide fast diagnosis of
unexpected behaviour; to allow the parameter space of the simulations
to be probed exhaustively; to improve the resolution of 
simulations at a fixed run-time, and to allow simulations to be run on
less powerful and cheaper hardware. 

In this paper we present a new approach that reduces the computational
time of simulations based on modal models by several orders of magnitude. 
We specifically target the computational cost of computing scattering matrices
for optical simulations. Our approach is based on a
near-optimal formulation of the integrals required to compute the scattering matrices, known
as a \textit{reduced order quadrature}~\cite{antil2012two} (ROQ). The reduced order quadrature has already
been applied in the context of astronomical data analysis with LIGO
\cite{PhysRevLett.114.071104} where the repeated computation of
quantities similar to the scattering matrix dominate the run time of
the analysis codes. Crucially, the reduced order quadrature is
designed to provide huge improvements to computational efficiency
whilst maintaining computational precision. 

The ROQ can be regarded as a type of near-optimal, application specific,
downsampling of the integrands needed to compute the integrals for the
scattering matrices \cite{antil2012two}. It is analogous to Gaussian
quadrature, but whereas Gaussian quadrature is designed to provide exact
results for polynomials of a certain degree, the ROQ produces nearly-exact
results for arbitrary parametric functions. Importantly, we are able to place
tight error bounds on the accuracy of the ROQ for a particular application~\cite{antil2012two}
making it an ideal technique to speed up costly integrals.
It exploits an offline/online methodology in which we recast the expensive
integrals used to compute scattering matrices into a more computationally
efficient form in the ``offline'' stage. This is then used for the rapid
``online'' evaluation of the scattering matrices. The offline stage can
itself be computationally expensive, however it need
only be performed once and is easily parallelised.
The data computed in the offline stage---that is needed by the
ROQ---can be stored and shared for particular use cases in
the online stage so that the offline cost does not need to be
factored in at run time.

We derive the algorithm in a general form and report on the implementation and performance of this
method in an example task for the LIGO interferometers. The
implementation of the method described in this article is available as 
open source as part of the \Finesse source code and the Python based
package \textsc{Pykat}~\cite{pykat_webpage}, which will also contain the
offline computed data to enable others to model Advanced LIGO like
arm cavities. Our particular implementation here is used to provide a
simple, real-word example. However, the algorithm can be easily
implemented in other types of simulation tools, for example, time
domain simulations or grid based tools (also known as FFT
simulations) that compare beam shapes. In all cases our algorithm can significantly reduce the
computation time for evaluating overlap integrals of Gaussian
modes with numerical data.
 
The paper is outlined as follows: In section~\ref{sec:hom} we give an overview
of the paraxial description of the optical eigenmodes and scattering into higher order modes. In Section~\ref{sec:int_interp} we provide
the mathematical background and algorithm for producing the ROQ. Section~\ref{sec:int_interp} heavily relies on an additional mathematical technique known as the ``empirical interpolation method'' \cite{Maday_2009}. We
assume no prior knowledge of this and provide the main details and results necessary for the ROQ.
Section~\ref{sec:example} then highlights an exemplary case to demonstrate our method
for modelling near-unstable optical cavities.
Finally in section~\ref{sec:performance} the computational performance of our method
is analysed.

\section{Higher-order optical modes}\label{sec:hom}

Gravitational wave detectors are constructed of multiple optical cavities
based on a Michelson interferometer.
The circulating laser beams in such an optical setup
is well described by the the paraxial Gaussian eigenmodes of
a spherical cavity; an efficient basis for describing the spatial properties of a
laser beam in the transverse plane to the propagation axis \cite{kogelnik65}.
The fundamental Gaussian mode is described in cartesian coordinates by:
\begin{multline}
u_{00}(x,y; q_x, q_y)  =  \sqrt{\frac{2}{\pi w_{x}(q_x) w_{y}(q_y)}}
		e^{-\I k \left( \frac{x^2}{2q_x} + \frac{y^2}{2q_y}\right)}
\end{multline}
$w_x$ and $w_y$ are the beam spot sizes in the $x$ and $y$ directions,
$k$ is the wavenumber of the laser light and $\mathbf{q} = \{q_x, q_y\}$ are the
complex \emph{beam parameters} in the $x$ and $y$ directions.
The shape of the Gaussian mode is fully defined by the wavelength of the
light $\lambda$ and the beam parameter:
\begin{equation}
q_x = z_x + \I z_{R,x} = z_x + \I \frac{\pi w_{0,x}^2}{\lambda}
\end{equation}
where $z_x$ is the distance from the waist, $z_{R,x}$ is the Rayleigh
range, $w_{0,x}$ is
the size of the waist and $\lambda = 1064$nm is the wavelength of the
Nd:YAG laser used in current GW detectors. The same set of parameters
exists for $q_y$.

Any perturbations in the beam's spatial profile from this 
fundamental Gaussian can be described by the
addition of higher-order Gaussian modes (HOMs).
In this paper we discuss in particular the cartesian orthogonal basis of
Hermite-Gaussian (HG) modes \cite{kogelnik65}, however our method 
is applicable for any
other suitable basis. The complex transverse spatial
amplitude of these HG modes is given by:
\begin{eqnarray}
&&u_{\mathrm{nm}}(x,y, q_x, q_y)=u_{\mathrm{n}}(x, q_x)u_{\mathrm{m}}(y, q_y) \nonumber \\
&&u_{\mathrm{n}}(x,q)=\left(\frac{2}{\pi}\right)^{1/4}\left(\frac{1}{2^nn!w_{0}}\right)^{1/2}
\left(\frac{q_{0}}{q}\right)^{1/2}\nonumber \\
&& \left(\frac{q_{0}~q^*}{q_{0}^*~q}\right)^{n/2}  H_n\left(\frac{\sqrt{2}x}{w(z)}\right) \exp\left(-\I\frac{kx^2}{2q}\right).
\label{eq:unm}
\end{eqnarray}
where $n$ defines the order of the Hermite polynomials $H_{n}$
in the $x$ axis and $m$ for the $y$.
The order of the optical mode is $\mathcal{O} = n+m$ and individual modes
are typically referred to as TEM$_{nm}$.
A laser field with a single optical frequency
component $\omega$ can be expanded into a beam basis whose shape is described by $\mathbf{q}$ as:
\begin{equation}
    E(x, y, t; \mathbf{q}) = \sum^{n+m \leq \maxtem}_{n=0,m=0} a_{nm} u_{nm}(x,y; \mathbf{q}) e^{\I\omega t}
\end{equation}
where $a_{nm}$ is a complex value describing the amplitude and phase 
of a mode TEM$_{nm}$ and \maxtem is the maximum order
of modes included in the expansion. 

\begin{figure*}[t]
      \includegraphics[width=1\textwidth]{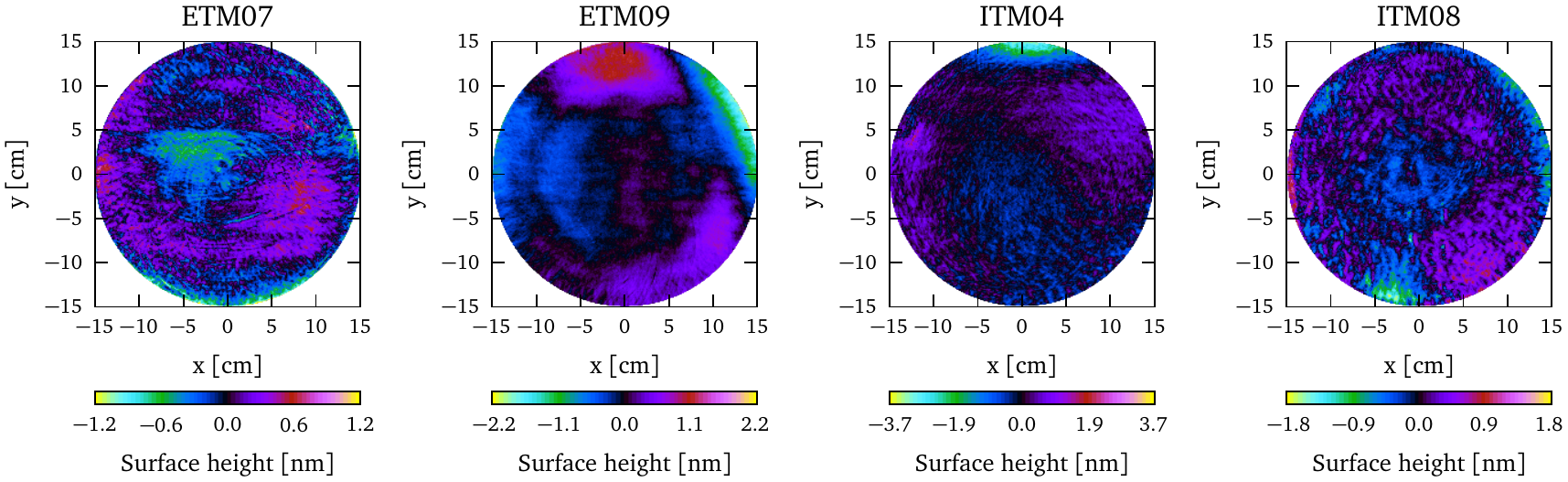}
        \caption{Measured surface distortions for the mirrors currently
		installed in the Livingston LIGO site (shown here are the
      distortions of the test masses before they were coated). ETM09 and ITM08 are installed
		in the x-arm and ETM07 and ITM04 in the y-arm~\cite{aligo_galaxy}. Theses
		measurements have been processed to remove the overall mirror curvatures, offset
		and tilts. }
        \label{fig:maps}
\end{figure*}

\subsection{Scattering into higher-order modes}
When a field interacts with an optical component its mode content is
typically changed.
Here we define scattering as the relationship between the mode content
of the outgoing beams $\bar{a}$ with a beam shape $\mathbf{q}$, and
the mode content
of the incoming beam $\bar{a}'$ described with $\mathbf{q}'$.
Mathematically this is simply $\bar{a} = \hat{k} \bar{a}'$ where $\hat{k}$ is known as
the \emph{scattering matrix}.
Now consider the spatial profile of a beam reflected from on an imperfect optic
$E'(x,y; \mathbf{q'}) = A(x,y)E_{\mathrm{in}}(x,y;\mathbf{q'})$, where $E_{\mathrm{in}}$ is the incident beam
and $A(x,y)$ is complex function describing the perturbation it has undergone.
For example, on reflection a beam will be clipped by the finite size of the mirror $\alpha(x,y)$
and reflected from a surface with height variations $z(x,y)$. Thus, both the amplitude and phase
of the beam will be affected and $A(x,y) = \alpha(x,y) e^{\I2k\, z(x,y)}$. An example of
the measured surface height variations present on LIGO test
mass mirrors can be seen in figure~\ref{fig:maps}~\cite{aligo_galaxy}. 
The mode content of the outgoing beam $E(x, y; \mathbf{q})$ is computed by projecting $E'$
into the outgoing beam basis $\mathbf{q}$.
For any incoming HOM $u_{n'm'}$  the amount of outgoing $u_{nm}$
can be computed via an overlap integral, this complex value is known as a \textit{coupling coefficient}:
\begin{multline}
k_{nm,n'm'}(q_x, q'_x, q_y, q'_y; A) = \\ 
\iint_{-\infty}^{\infty}\mathcal{K}(\boldsymbol{\lambda}_x;x) A(x,y) 
 \mathcal{K}(\boldsymbol{\lambda}_y;y)\, dx\,dy\,,
\label{eq:knm}
\end{multline}
where the integral kernels $\mathcal{K}(\boldsymbol{\lambda}_x;x)$ and $\mathcal{K}(\boldsymbol{\lambda}_y;y)$ are given by
\begin{eqnarray}
\label{eq:kernels}
\mathcal{K}(\boldsymbol{\lambda}_x;x) &=& u^{*}_n(x, q_x)u_{n^{\prime}}(x, q^{\prime}_x)\,,\\
\mathcal{K}(\boldsymbol{\lambda}_y;y) &=& u^{*}_m(y, q_y)u_{m^{\prime}}(y, q^{\prime}_y)\,,
\end{eqnarray}
and the parameter vectors are given by $\boldsymbol{\lambda}_x = (n, n^{\prime}, q_x, q'_x)$ and
$\boldsymbol{\lambda}_y= (m, m^{\prime}, q_y, q'_y)$.
There are two general cases when computing \eqref{eq:knm}: $q \neq
q'$ which we refer to as \textit{mode-mismatched} and $q =
q'$ as \textit{mode-matched}.

Computing the scattering matrix $\hat{k}$ requires evaluating the
integral~\eqref{eq:knm} for each of its elements.
If couplings between modes up to and including order $\mathcal{O}$
are considered then the number of elements in
$\hat{k}$ is $N_k(\mathcal{O}) = (\mathcal{O}^4 + 6\mathcal{O}^3 +
13\mathcal{O}^2+ 12\mathcal{O} +4)/4$ and the computational cost of
evaluating this many integrals can be very expensive.
In our experience \cite{T1300954, czbthesis}
a typical LIGO simulation
task involving HOMs can be performed with $\mathcal{O}=6-10$
while some cases, such as those that include strong thermal distortions or clipping,
a higher maximum order is required.

In simple cases where $A(x,y)=1$ or $A(x,y)$ represents a tilted
surface, analytical
results are available for both mode matched and mismatched
cases~\cite{bayer-helms, VBP2}. 
In general however
$A(x,y)$ is of no particular form and the integral in \eqref{eq:knm}
must be evaluated numerically.
It is possible to split multiple distortions into separate scattering matrices
$A(x,y) \Rightarrow A(x,y)\,B(x,y)$ and
the coupling coefficients become a product of two separate matrices:
\begin{multline}
k_{nm,n'm'}(\mathbf{q}, \mathbf{q}'; A\,B) = 
\\\sum_{\tilde{n}, \tilde{m} = 0}^{\infty}
k_{nm,\tilde{n}\tilde{m}}(\mathbf{q}, \tilde{\mathbf{q}}; A) \ 
k_{\tilde{n}\tilde{m},n'm'}(\tilde{\mathbf{q}}, \mathbf{q}'; B)
\end{multline}
where $\tilde{\mathbf{q}}$ is an expansion beam parameter which we are free to choose.
Thus our scattering matrix becomes $\hat{k}(\mathbf{q},\mathbf{q}') = \hat{k}_A(\mathbf{q}, \tilde{\mathbf{q}}) \hat{k}_B(\tilde{\mathbf{q}},\mathbf{q}')$.
By choosing $\tilde{\mathbf{q}} = \mathbf{q}$ or $\mathbf{q}'$ we can
set the mode-mismatching to be in either one matrix or the other. This is
ideal as a mode-matched $\hat{k}$ is a Hermitian matrix whose 
symmetry can be exploited to only compute one half of the matrix.
By ensuring that this matrix also contains any distortions that require
numerical integration the computational cost can be nearly halved.
It is then possible to benefit from the fast analytic solutions to \eqref{eq:knm}
to account for mode-mismatching in the other matrix.


\section{Efficiently Computing Scattering Matrices: Integration by Interpolation}\label{sec:int_interp}

For a discretely sampled mirror map with $L$ sample points in both the $x$ and
$y$ directions, the coupling coefficient \eqref{eq:knm} can be approximated using a
composite Newton-Cotes quadrature rule:
\begin{multline}
k_{nm,n'm'}(q_x, q'_x, q_y, q'_y; A) \\
\approx \sum_{k=1}^L \sum_{l=1}^L W_{kl}  \mathcal{K}(\boldsymbol{\lambda}_x;x_k)\,
A(x_k, y_l) \mathcal{K}(\boldsymbol{\lambda}_y;y_l),
\label{eq:discteteknm} 
\end{multline}
where $W$ is an $L\times L$ matrix describing the 2D composite Newton-Cotes
quadrature weights over the area of the map. The matrix is found by taking the outer product
of the 1D composite Newton-Cotes quadrature weights~\cite{ralston2001first} in both $x$ and $y$ directions.
There are $L^{2}$ terms in the double sum
\eqref{eq:discteteknm}. When $L^2$ is large, as in the cases of
interest for this paper, there are two major bottlenecks: $(i)$
evaluation of the kernel at each discrete $x_k, y_k$ and, $(ii)$
evaluation of the double sum.
With a set of $M \ll L$ basis elements that accurately spans the kernel
space, it is possible to replace the double sum \eqref{eq:discteteknm}
with a \emph{reduced order quadrature} (ROQ) rule \eqref{eq:roq} 
containing only $M^{2}$ terms, reducing the overall cost of the by a
factor of $\sim L^2/M^{2}$, provided the kernel can be directly evaluated.

The reduced order quadrature scheme is implemented in three steps. The
first two are carried out offline, while the final,
mirror-map-dependent step is performed in preparation for the
simulations; once per map. The steps are as follows:
\textit{Step 1} - Construct a reduced basis (offline); a set of $M$ basis elements whose span describes the
kernel space. \textit{Step 2} - Construct an interpolant using the basis (offline)
by requiring it to exactly match any kernel at  
$M$ carefully chosen spatial subsamples $\{X_k\}_{k=1}^M$ \cite{Barrault2004667}
(and similarly for $y$).
\textit{Step 3} - Use the interpolant to replace the inner product
evaluations in ~\eqref{eq:discteteknm} with the ROQ~\eqref{eq:roq} (online). 

\subsection{The Empirical Interpolation Method}\label{sec:eim}

The empirical interpolation method is an efficient technique performing this
offline/online procedure and has been demonstrated in the context of
astronomical data analysis with LIGO \cite{PhysRevLett.114.071104}. Provided the kernels vary
smoothly with $\boldsymbol{\lambda}_x$ over $x$ and $\boldsymbol{\lambda}_y$ over $y$ then there
exists a set of kernels at judiciously chosen parameter values that
represent any kernel - and hence any integral \eqref{eq:knm} - for an
arbitrary parameter value. This set of kernels constitutes the reduced
basis: Given any parameter value $\boldsymbol{\lambda}_x$ or $\boldsymbol{\lambda}_y$ we can find
the best approximation to the kernel at $\boldsymbol{\lambda}_x$ or $\boldsymbol{\lambda}_y$
as linear combination of the reduced basis. 

The ability to exploit the reduced basis to quickly evaluate \eqref{eq:knm}
depends on being able to find an affine parameterization of the
integral kernels. In general, the kernels do not admit such a
parameterization. However, the empirical interpolation method finds a near-optimal affine
approximation whose accuracy is bounded by the accuracy of the reduced basis
\cite{antil2012two}. This affine approximation is called the \emph{empirical interpolant}. The spatial integrals
over $dx\,dy$ in \eqref{eq:knm} will only depend on the reduced basis (and hence only have to be computed once for a
given mirror map) and the parameter variation is handled by the empirical interpolant at a reduced computational cost. 

The empirical interpolation method exploits the offline/online computational concept where we
decompose the problem into a (possibly very) expensive offline part
which affords a cheap online part. In this case, the expensive offline
part is in finding the reduced basis and constructing the empirical interpolant. Once the empirical interpolant is
found then we use it for the fast online evaluation of \eqref{eq:knm}. 
One of the main reasons why the empirical interpolant is used for fast online evaluation
of \eqref{eq:knm} is due to its desirable error properties that makes
it superior to other interpolation methods, such as polynomial
interpolation. In addition, the empirical interpolant avoids many of the pitfalls of
high-dimensional interpolation that we would otherwise encounter (see,
e.g. \cite{MZA:9150038}).

\subsection{Affine Parameterization} 
We would like the kernel to be separable in the mode parameters $(\boldsymbol{\lambda}_x, \boldsymbol{\lambda}_y)$ and spatial
position $(x,y)$. For these reasons we will look for a representation
of the kernel that has the following form:
\begin{eqnarray}\label{eq:objfn}
\mathcal{K}(\boldsymbol{\lambda}_x;x) &=& a(\boldsymbol{\lambda}_x)\,f(x)\,,\nonumber \\ 
\mathcal{K}(\boldsymbol{\lambda}_y;y) &=& a(\boldsymbol{\lambda}_y)\,f(y)\,.
\end{eqnarray}
The functions $a$ and $f$ are
the same irrespective of whether the kernel is a function of $x$ or
$y$ due to the symmetries of the Hermite Gauss modes. Using the affine
parameterization, the coupling coefficient \eqref{eq:knm} is:
\begin{multline}
k_{nm,n'm'}(q_x, q'_x, q_y, q'_y) = a^{*}(\boldsymbol{\lambda}_x)a(\boldsymbol{\lambda}_y)\\\iint_{-\infty}^{\infty}{ f^{*}(x) A(x,y) f(y)} dx\,dy\,,
\end{multline}
This affine parameterization thus allows us to compute all the parameter-dependent pieces
efficiently in the online procedure as all the $x-y$ integrals are performed only once for a given mirror map.
In general the kernel will not admit an exact affine decomposition as in \eqref{eq:objfn}.
Using the EIM, the approximation to the kernels will have the form:
\begin{eqnarray}
\label{eq:EIexpansion}
\mathcal{K}(\boldsymbol{\lambda}_x;x) &\approx& \sum_i\,c_i(\boldsymbol{\lambda}_x)\,e_i(x)\,,\\
\mathcal{K}(\boldsymbol{\lambda}_y;y) &\approx& \sum_i\,c_i(\boldsymbol{\lambda}_y)\,e_i(y)\,\nonumber.
\end{eqnarray}
The sum is over the reduced basis elements
$e_i$ and coefficients $c_i$ that contain the parameter dependence. 

Given a basis $e_i(x)$, the $c_i(\boldsymbol{\lambda}_x)$ in \eqref{eq:EIexpansion} are the solutions to the M-point interpolation
problem whereby we require the interpolant to be exactly equal to the kernel at any parameter value $\boldsymbol{\lambda}_x$ at a
set of interpolation nodes $\{X\}_{i=1}^{M}$:
\begin{eqnarray}
 \label{eq:interp_problem} 
\mathcal{K}(\boldsymbol{\lambda}_x;X_j) &=& \sum_{i=1}^M  c_i(\boldsymbol{\lambda}_x) e_i(X_j) = \sum_{i=1}^{M}V_{ji}\,c_i(\boldsymbol{\lambda}_x),
\end{eqnarray}
where the matrix $V$ is given by
\begin{equation} \label{eq:InterpMatrix}
  V \equiv \left(  \begin{array}{cccc}   
              e^1(X_1)  &  e^2(X_1)            & \cdots & e^{M}(X_1)      \\
              e^1(X_2)  &  e^2(X_2)            & \cdots & e^{M}(X_2)       \\
              e^1(X_3)  &  e^2(X_3)          & \cdots & e^{M}(X_3)   \\              
              \vdots    & \vdots             & \ddots & \vdots                       \\
              e^1(X_{M})  & e^2(X_{M})    & \cdots & e^{M}(X_{M})  \\               
             \end{array}
   \right) 
\end{equation} 
Thus we have:
\be
c_i(\boldsymbol{\lambda}_x) = \sum_{j=1}^M \left( V^{-1}\right)_{ij}\,\mathcal{K}(\boldsymbol{\lambda}_x;X_j)\,.  \label{eq:CEIM}
\ee

Substituting \eqref{eq:CEIM} into \eqref{eq:EIexpansion}, the empirical interpolant is:
\be \label{eq:EIM_with_B}
\mathcal{I}_{M}[\mathcal{K}](\boldsymbol{\lambda}_x;x) = \sum_{j=1}^M \mathcal{K}(\boldsymbol{\lambda}_x;X_j) B_j (x) 
\ee 
where:
\be
B_j(x) \equiv  \sum_{i=1}^M e_i (x )  \left( V^{-1} \right)_{ij}    \label{eq:BEIM}
\ee
and is independent of $\boldsymbol{\lambda}_x$.
The special spatial points $\{X_k\}_{k=1}^M$, selected from a discrete set of points along
$x$, as well as the basis can be found using
 Alg.~(\ref{alg:Greedy}) which is described in the next section. 
 
 We note that the kernels $\mathcal{K}(\boldsymbol{\lambda}_x;x)$ appear explicitly on the right hand side of
 \eqref{eq:EIM_with_B}. Because of this, we have to be able to directly evaluate the
 kernel at the empirical interpolation nodes $\{X_k\}_{k=1}^M$. Fortunately this is
 possible in this case as we have closed form expressions for the kernels. If the
 kernels were solutions to ordinary or partial differential equations that needed
 to be evaluated numerically then using the empirical interpolant becomes more
 challenging, however this is not required here (see, e.g., \cite{Gunzburger20071030, Barrault2004667, libMeshPaper} for applications of the empirical interpolation method to ordinary and partial differential equation solvers).

\subsection{The Empirical Interpolation Method Algorithm (Offline)}

The empirical interpolation method algorithm solves \eqref{eq:EIM_with_B} for arbitrary $\boldsymbol{\lambda}_x$. While it would be
possible in principle to use arbitrary basis functions, such as Lagrange polynomials which are
common in interpolation problems \cite{Press:2007:NRE:1403886, citeulike:3049920}, we take a different approach that uses only
the information contained in the kernels themselves. We will take
as our basis a set of $M$ judiciously chosen kernels sampled at points on the parameter space $\{\boldsymbol{\lambda}_x^i\}_{i=1}^{M}$, where $M$ is equal to the number of basis elements in \eqref{eq:EIM_with_B}. Because the kernels
vary smoothly with $\boldsymbol{\lambda}_x$ a linear combination of the basis elements will give a
good approximation to $\mathcal{K}(\boldsymbol{\lambda}_x;x)$ for any parameter value \cite{Barrault2004667}. We can
then build an interpolant using this basis by matching $\mathcal{K}(\boldsymbol{\lambda}_x;x)$ to the
span of the basis at a set of $M$ interpolation nodes $\{X_k\}_{k=1}^M$. The empirical interpolation method algorithm, shown
in Alg.~(\ref{alg:Greedy}), provides both the basis and the nodes. 

The empirical interpolation method algorithm uses a greedy procedure to select the reduced basis elements and interpolation nodes. With
the greedy algorithm, the basis and interpolant are constructed iterative whereby the interpolant on each iteration is optimized according to an appropriate error measure. This guarantees that the error of the interpolant is on average decreasing and - as we show in section~\ref{sec:errors} - that the interpolation error decreases exponentially quickly. We follow Algorithm
3.1 of \cite{aanonsen2009empirical} which is reproduced in Alg.~(\ref{alg:Greedy}). 
 
The first input to the algorithm is a \emph{training space} (TS) of kernels - distributed on the parameter
space $\boldsymbol{\lambda}_x$ - and the associated set of parameters. This training space
is denoted by $ \mathcal{T} = \{ \boldsymbol{\lambda}^k_x \, , \mathcal{K}(\boldsymbol{\lambda}^k_x;x) \}_{i=1}^N$
and should be densely populated enough to represent the full space of kernels as faithfully as possible. Hence it is important that
$1 \ll N$. The second input is the desired maximum error of the interpolant $\epsilon$. We find that the
$L^{\infty}$ norm is a robust error measure for the empirical interpolant and hence $\epsilon$
corresponds to the largest tolerable difference between the empirical interpolant and any kernel in the
training set $\mathcal{T}$.

The algorithm is initialized on steps 3 and 4 by setting the zeroth order interpolant
to be zero, and defining the zeroth order interpolation error to be infinite. The greedy algorithm proceeds as follows:
We identify the basis element on iteration $i$ to be the $\mathcal{K}(\boldsymbol{\lambda}_x;x) \in \mathcal{T}$ that maximizes the $L^{\infty}$ norm with the
interpolant from the previous iteration, $\mathcal{I}_{i-1}[\mathcal{K}](\boldsymbol{\lambda}_x;x)$. This is performed in
steps 7 and 8.
On step 9 we select $X_i$, the $i^{th}$ interpolation node, by selecting the position at
which the largest error occurs, and adding that position to the set of interpolation nodes. By definition, the interpolant
is equal to the underlying function at the interpolation nodes and so the error at $X_i$ - which is the largest error on the current iteration - is removed.
On step 10 we normalize the basis function. This ensures that the matrix \eqref{eq:InterpMatrix} is well
conditioned. On steps 11 and 12 we compute \eqref{eq:InterpMatrix} and \eqref{eq:BEIM},
which are used to construct the empirical interpolant \eqref{eq:EIM_with_B}. Finally, on step 13
we compute the interpolation error $\sigma_i$ between the interpolant on the current iteration
$\mathcal{I}_{i}[\mathcal{K}](\boldsymbol{\lambda}_x;x)$ and $\mathcal{K}(\boldsymbol{\lambda}_x;x) \in \mathcal{T}$
as in step 7. The procedure is repeated until $\sigma_i \leq \epsilon$. 

Once the interpolant for $\mathcal{K}(\boldsymbol{\lambda}_x;x)$ is found, the equivalent interpolant for $\mathcal{K}(\boldsymbol{\lambda}_y;y)$ is obtained trivially from $\mathcal{I}_{M}[\mathcal{K}](\boldsymbol{\lambda}_x;x)$ by setting $x \rightarrow y$.

{%
\begin{algorithm}[H]
\caption{
\small
Empirical Interpolation Method Algorithm:
The empirical interpolation method algorithm builds an interpolant for the kernels \eqref{eq:kernels} iteratively using a greedy procedure. On each iteration the current interpolant is validated against a ``training set'' $\mathcal{T}$ of kernels and the worst interpolation error is identified. The interpolant is then updated so that it describes the worst-error point perfectly. This is repeated until the worst error is less than or equal to a user specified tolerance $\epsilon$.}
\label{alg:Greedy}
\begin{algorithmic}[1]
\small
\State {\bf Input:} $ \mathcal{T} = \{ \boldsymbol{\lambda}^k_x \, , \mathcal{K}(\boldsymbol{\lambda}^k_x;x) \}_{k=1}^N$ and $\epsilon$ 
\vskip 10pt
\State Set $i=0$
\vskip 3pt
\State Set $\mathcal{I}_0[\mathcal{K}](\boldsymbol{\lambda}_x;x) = 0$
\vskip 3pt
\State Set $\sigma_0 = \infty$
\vskip 3pt
\While{$\sigma_i \ge \epsilon$}
\vskip 3pt
\State $i \rightarrow i+1$
 \vskip 3pt
\State $\boldsymbol{\lambda}_x^i = \arg \underset{\boldsymbol{\lambda}_x \in \mathcal{T}}{\max}\vert\vert\mathcal{K}(\boldsymbol{\lambda}_x;x) - \mathcal{I}_{i-1}[\mathcal{K}](\boldsymbol{\lambda}_x;x)\vert\vert_{L^{\infty}}$
\vskip 3pt
\State $\xi_i(x) = \mathcal{K}(\boldsymbol{\lambda}^i_x;x)$
\vskip 3pt
\State $X_i = \arg\underset{x}{\max}|\xi_i(x) - \mathcal{I}_{i-1}[\xi_i](x)|$
\vskip 3pt
\State $e_i(x) = \frac{\xi_i(x) - \mathcal{I}_{i-1}[\xi_i](x)}{\xi_i(X_i) - \mathcal{I}_{i-1}[\xi_i](X_i)}$
\vskip 3pt
\State $V_{lm} = e_l(X_m)\,\, l \leq i, m \leq i$
\vskip 3pt
\State $B_m(x) = \sum_{l} e_l (x )  \left( V^{-1} \right)_{lm} l \leq i, m \leq i$
\vskip 3pt
\State $\sigma_i = \underset{\boldsymbol{\lambda}_x \in \mathcal{T}}{\max}\vert\vert\mathcal{K}(\boldsymbol{\lambda}_x;x) - \mathcal{I}_{i}[\mathcal{K}](\boldsymbol{\lambda}_x;x)\vert\vert_{L^{\infty}}$
\EndWhile
\vskip 10pt
\State {\bf Output:} Interpolation matrix $\{B_j(x)\}_{j=1}^M$ and interpolation nodes $\{ X_j \}_{j=1}^M$.
The equivalent interpolant for $\mathcal{K}(\boldsymbol{\lambda}_y;y)$ is obtained trivially from $\{B_j(x)\}_{j=1}^M$ and $\{ X_j \}_{j=1}^M$ by setting $x\rightarrow y$ and $X \rightarrow Y$.
\end{algorithmic}
\end{algorithm}
}

\subsection{Error Bounds on the Empirical Interpolant}\label{sec:errors}

Before we proceed to demonstrate the utility of the empirical interpolant for quickly
evaluating \eqref{eq:knm} we briefly remark on some of the error properties of the empirical interpolation method.
A more detailed error analysis of the empirical interpolant can be found in \cite{antil2012two}.
For our purposes the empirical interpolant possess a highly desirable property, namely
exponential convergence to the desired accuracy $\epsilon$. It can be shown \cite{aanonsen2009empirical, Maday_2009}
(though we do not do so here) that there exists constants $c > 0$ and $\alpha > \log(4)$
such that for any function $f$ the empirical interpolant satisfies 
\begin{equation}
\label{eq:error_bound}
|| f - \mathcal{I}_{M}[f]||_{L^{\infty}} \leq c\,e^{-(\alpha - \log(4))M}\,.
\end{equation} 
This states that under the reasonable assumption that there exists an order $M$ interpolant that allows for exponential convergence, then the empirical interpolation method will ensure that we converge to this interpolant exponentially quickly. This is an important property as it means that the order of the interpolant, $M$, tends to be small for practical purposes.
In addition, because the quantity on the right hand side $c\,e^{-(\alpha - \log(4))M}$ is set to a user specified tolerance $\epsilon$ then we can set an \textit{a priori} upper bound on the \textit{worst}-fit of the interpolant. However, one must still verify that the interpolant describes functions outside the training \textit{a postiori}, though the error bound should still be satisfied provided that the training set was dense enough. In fact, it can be shown \cite{Maday_2009} that the empirical interpolation method is a near optimal solution to the Kolmogorov $n$-width problem in which one seeks to find the best $M$-dimensional (linear) approximation to a space of functions.

It is important to recall that in this paper we are interpolating the integral kernels \eqref{eq:kernels} which are a function of six free parameters $\boldsymbol{\lambda}_x$: two indices $n$ and $n^{\prime}$ and two complex beam parameters $q_x$ and $q_x^{\prime}$. Had we not used the EIM, we would have had to find an alternative way of expressing the $\boldsymbol{\lambda}_x$-dependent coefficients in \eqref{eq:EIexpansion}. Consider, for example, a case in which we had used tensor-product splines to describe the coefficients: Using a grid of just ten points in each of the six parameters in $\boldsymbol{\lambda}_x$ would result in an order $10^6$ spline which would surely be computationally expensive to evaluate. Furthermore, there would be no guarantee of its accuracy or convergence to a desired accuracy.

\begin{figure*}[bht]
    \centering
    \begin{subfigure}[b]{0.49\textwidth}
        \centering
        \includegraphics[width=\textwidth]{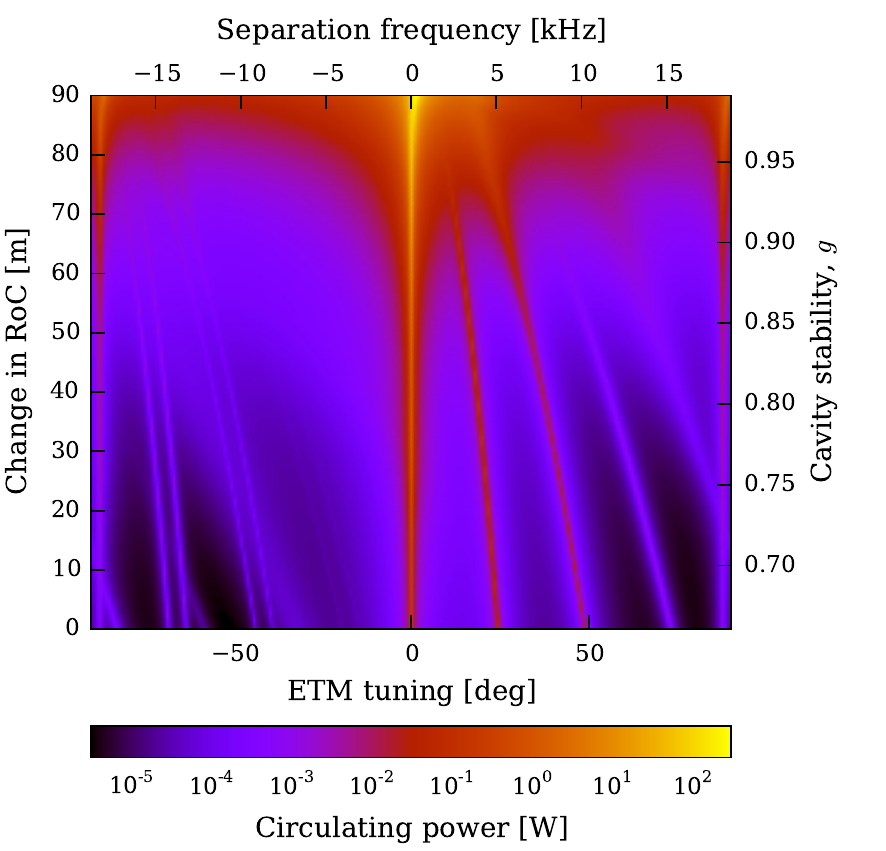}
        \caption{Cavity scan as ITM and ETM RoC varied}
        \label{fig:aligo_cav_instable_scan}
    \end{subfigure}
    \hfill
    \begin{subfigure}[b]{0.49\textwidth}
        \centering
        \includegraphics[width=\textwidth]{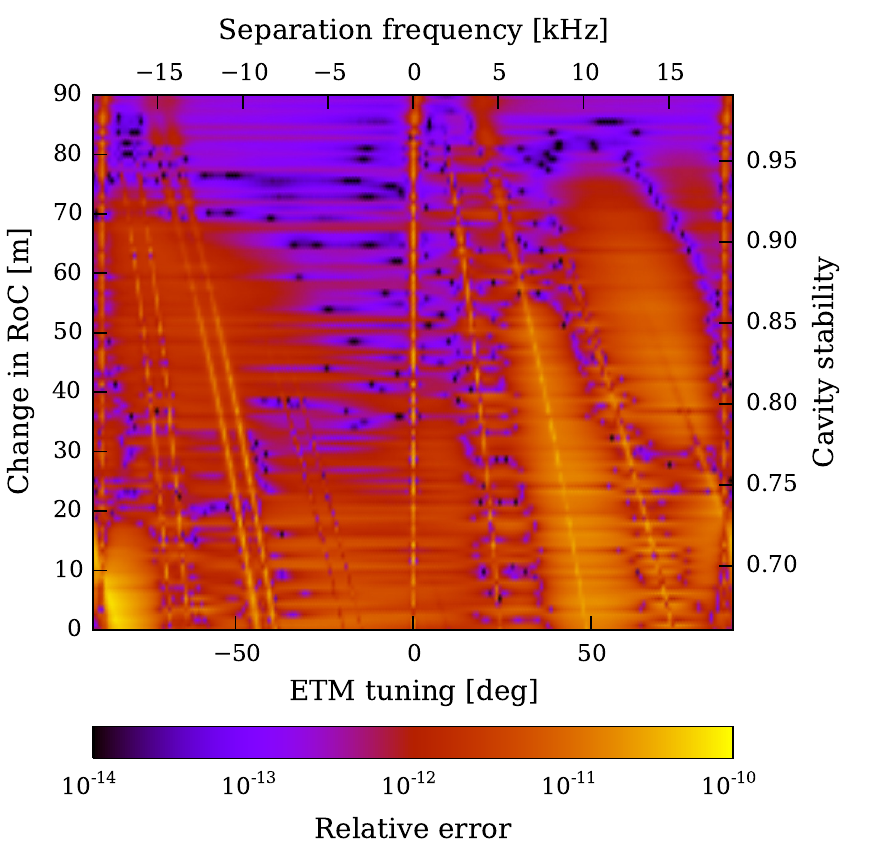}
          \caption{The relative error between ROQ and Newton-Cotes.}
          \label{fig:aligo_cav_instable_scan_err}
    \end{subfigure}
	
    \caption{Modeled LIGO cavity scan as the RoC of the
	ITM and ETM are varied to make the cavity increasingly more
  unstable. This simulation was run for $\maxtem=10$ and includes
  clipping from the finite size of the mirrors and surface
  imperfections from the ETM08 and ITM04 maps.
  Figure~\ref{fig:aligo_cav_instable_scan} shows how 
  the amount of power scattering into HOM changes as $g
  \rightarrow 1$. Also visible here is the reduction in the mode
  separation frequency with increasing instability. The contribution
  of the TEM$_{00}$ mode  has been removed to make the HOM
  content more visible. The reduced basis was built for mode order
  $\mathcal{O}=14$, to reduce errors, see 
  figure~\ref{fig:param_range_err}. 
  The difference in this result when using ROQ compared to
  Newton-Cotes is shown in \ref{fig:aligo_cav_instable_scan_err}.}
    \label{fig:aligo_cav_instable}
\end{figure*}

\subsection{Reduced order quadrature (Online)}

Substituting the empirical interpolant
\eqref{eq:EIM_with_B} into \eqref{eq:discteteknm} gives the ROQ,
\begin{multline}
k_{nm,n^{\prime}m^{\prime}}(q_x, q'_x, q_y, q'_y; A) = \\ \sum_{k=1}^M \sum_{l=1}^M w_{kl}\,\mathcal{K}(\boldsymbol{\lambda}_x;X_k)\ \mathcal{K}(\boldsymbol{\lambda}_y;Y_l) \, ,
\label{eq:roq}
\end{multline}
with the ROQ weights $\omega_{kl}$ given by:
\begin{equation}
\omega_{kl} = \sum_{i=1}^L\, \sum_{j=1}^L W_{ij} A(x_i, y_j) B_k (x_i)\,B_l (y_j)  \, .
\label{eq:wk}
\end{equation}
The ROQ form of the coupling coefficient enables fast online
evaluations of the coupling coefficients. Note that because only $M^2$
operations are required to perform the double sum \eqref{eq:roq} we
expect that the ROQ is faster than the traditional $L^2$-term
Newton-Cotes integration by a factor of $L^2/M^2$ provided that $M <
L$. We expect in practice that $M \ll L$ due to the exponential convergence of the empirical interpolation method.

\begin{figure}
      \includegraphics[width=0.45\textwidth]{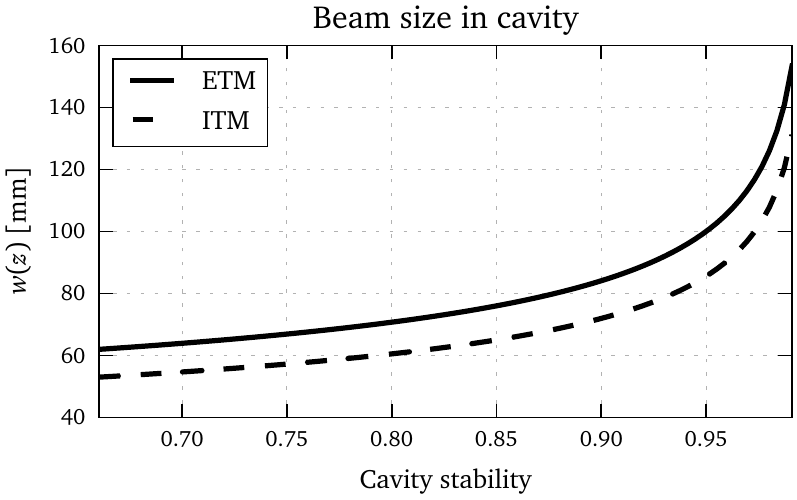}
      \caption{The beam size on the ITM and ETM of a LIGO cavity as a
        function of cavity stability parameter as the mirror RoCs are tuned.}
      \label{fig:aligo_cav_instable_wz}
\end{figure}

The number of operations in \eqref{eq:roq} can be compressed further still due to the separability of the empirical interpolant
\eqref{eq:EIM_with_B} into beam parameters $\boldsymbol{\lambda}_x$ and
spatial position $x$ that allows us to exploit the spatial symmetry in the HG modes. The HG modes exhibit spatial symmetry/antisymmetry
under reflection about the origin. Hence it is useful to split the $x$
and $y$ dimensions into four equally sized quadrants and perform the
ROQ in each quadrant separately. 
For example, when a HG mode is symmetric between two or four of the 
quadrants then only two or one set(s) of
coefficients $\{\mathcal{K}(\boldsymbol{\lambda}_x;X_k)\}_{k=1}^M$ needs to be computed
(and likewise for $\{\mathcal{K}(\boldsymbol{\lambda}_y;Y_l)\}_{l=1}^M$). This will
speed up the computation of the ROQ~\eqref{eq:roq} by up to a factor
of four. Hence, in practice we need only build the
EI over one half plane for either positive or negative
values of $x$ (or equivalently $y$); we derive the basis spanning the second
half-plane by reflecting the basis about the origin.
To ensure that this symmetry is exploitable the data points of the
map must be distributed equally and symmetrically about the beam axis
($(x,y)=(0,0)$). Those points that lie on the $x$ and $y$ axes must also be
weighted to take into account they contribute to multiple quadrants
when the final sum is computed.
In the cases where the map data points are not correctly aligned
we found that bilinear interpolation of the data to retrieve 
symmetric points did not introduce any significant errors. However, higher-order interpolation methods can introduce
artefacts to the map data.

\section{Exemplary case: Near-unstable cavities and control signals} \label{sec:example}

There are several scenarios when modelling tools can benefit 
heavily from the ROQ method, of particular interest are cases where
the simulation time is dominated
by the integration time of the mirror surface maps.
One such example is an investigation into the feasibility of upgrading
the LIGO interferometers with near-unstable arm cavities.
The stability of a Fabry-Perot cavity is determined
by its length $L$ and radius of curvature (RoC) of each of its mirrors
and can be described using the parameter::
\begin{equation}
g = (1-L/Rc_{\mathrm{itm}})(1-L/Rc_{\mathrm{etm}}).
\label{eq:stability}
\end{equation}
with $0\leq g \leq 1$ defining the stable region.
Near-unstable cavities are of interest because they result in larger beam sizes
on the cavity mirrors (see also figure~\ref{fig:aligo_cav_instable_wz}) which reduces the
coating thermal noise~\cite{vinet09}, one of the limiting noise sources of the detector.
One negative aspect of such near-unstable cavities is that the
transverse optical mode separation frequency approaches zero as $g
\rightarrow 0$ or $1$.
The mode separation frequency determines the difference in resonance frequency
of higher-order modes with respect to the fundamental mode.
Thus with a lower separation frequency any defect in the cavity causing scattering
into HOMs is suppressed less and can contaminate control signals
for that cavity and couple extra noise into the GW detectors
output \footnote{Another potential problem is additional clipping or scattering
of the beam on the mirrors due to the larger beam sizes 
which can result in increased roundtrip losses of the arm cavity.}. 
The optimal cavity design must be determined as a trade-off between
these degrading effects and the reduction in coating thermal noise. This
is a typical task where a numerical model can be employed to search
the parameter space. In this case each point in that parameter space
corresponds to a different beam size in the cavity which forces a
re-computation of the scattering matrices on the mirrors. Thus the new
algorithm described in this paper should yield a significant
reduction in computing time.

In this section we briefly summarise the results from the
simulations and in the following section we provide the details of
setting up the model  and give an analysis of the performance of the
ROQ algorithm. We have implemented the ROQ integration in our
open-source simulation tool \Finesse and use the official input
parameter files for the LIGO detectors~\cite{aligo_file}.
Below we show the preliminary investigation of the behaviour of
a single Advanced LIGO like arm cavity with a finesse of 450, where
the mirror maps for the mirrors ETM08 and ITM04
\footnote{The nominal radius of curvatures of ETM08 and ITM04 are
$1934$m and $2245$m respectively. The optical properties of
these mirrors were taken from~\cite{aligo_galaxy}.}
were applied to the high reflective (HR) surfaces.
Note that we do not report the scientific results of the simulation
task which will be published elsewhere. This example is representative
for a class of modelling performed regularly for the LIGO
commissioning and design and provides us with a concrete and
quantitive setup to demonstrate the required steps to use the
ROQ algorithm.

\begin{figure}
    \centering
    \includegraphics[width=0.45\textwidth]{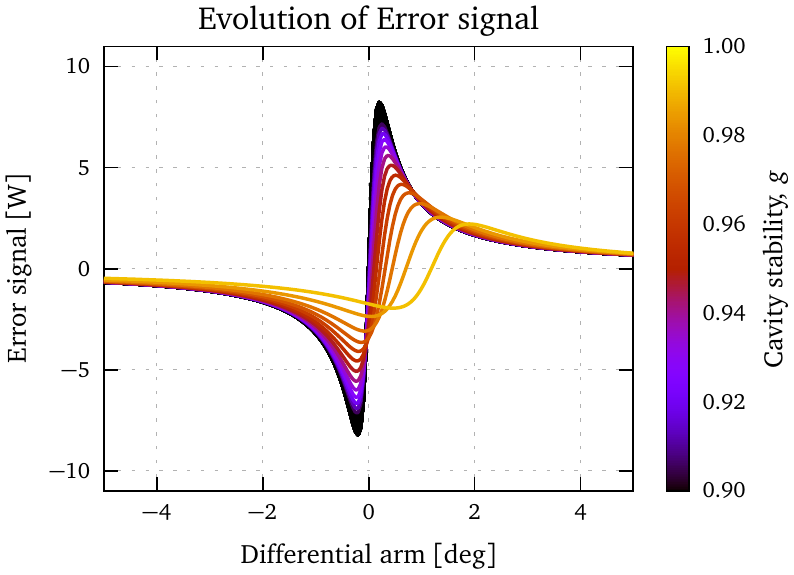}
    \caption{The Pound-Drever-Hall error signal for the LIGO cavity
      modelled in figure~\ref{fig:aligo_cav_instable}. A significant
      change in zero-crossing position and shape can be seen as the
      stability of the cavity is reduced ($g\rightarrow 1$).}
    \label{fig:aligo_cav_instable_errsig}
\end{figure}

\begin{figure*}[htb]
      \includegraphics[width=1\textwidth]{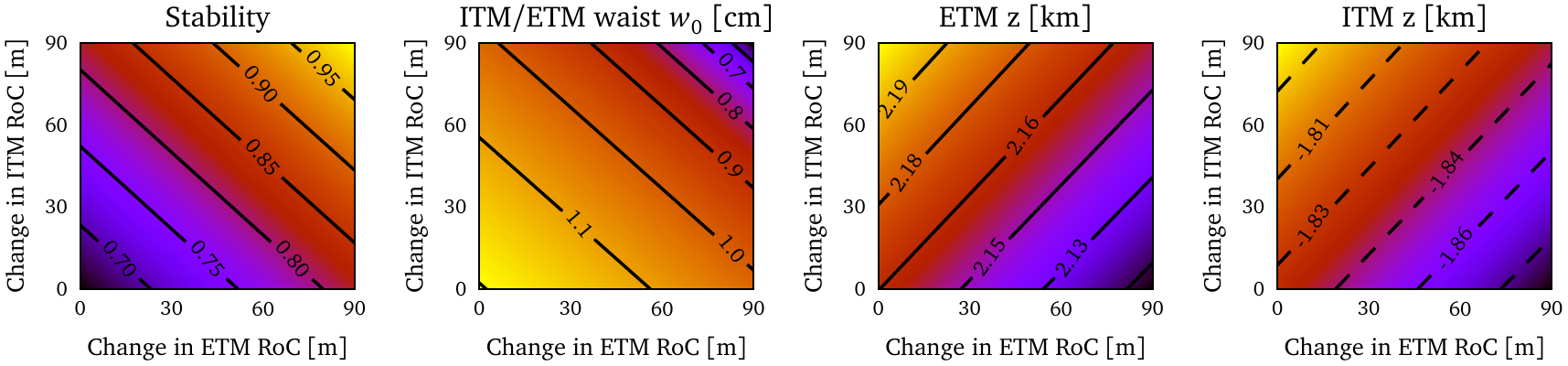}
        \caption{Range of beam parameters
		needed to model a change in curvature from 0\,m to 90\,m at the ITM
    and the ETM. In order to utilise the ROQ 
    to cover this parameter space, the empirical interpolant needs to be constructed using a
    training set made from kernels \eqref{eq:kernels} densely
    covering this space.}
        \label{fig:param_range}
\end{figure*}

Modelling the LIGO cavity for differing stabilities involves varying
the RoC of both the ITM and ETM. The resulting change in $w(z)$ at
each surface means the scattering matrices will need to be recomputed
for each state we choose. To view the HOM content in the cavity
created by the scattering a \textit{cavity scan} can be
performed, displacing one of the cavity mirrors along
the cavity axis on the order of the wavelength of the laser light,
$\lambda = 1064$\,nm, to change the resonance condition of the
cavity. 
We have performed the simulations using
$\mathcal{O}=10$ with Newton-Cotes integration and our ROQ method.
The results for cavity scans at different RoCs are shown in
figure~\ref{fig:aligo_cav_instable_scan}.	
The dominant mode is the fundamental
TEM$_{00}$ whose resonance defines the zero tuning, the power
in the TEM$_{00}$ mode has been removed from this plot to better show the
lower power HOM content. For more stable cavities (at the bottom of
the plot in figure~\ref{fig:aligo_cav_instable_scan})	the HOMs are well
separated and not resonant at the same time as the TEM$_{00}$. As the
RoC is increased, the stability is reduced and the HOMs can be
seen to converge and eventually become
resonant at a tuning of 0. At a stability of $g\approx 0.98$ the
cavity mode begins to break down significantly and many modes 
become resonant. The effect of this on a sensing and control 
signal used for a Pound-Drever-Hall control system is shown in
figure~\ref{fig:aligo_cav_instable_errsig}, where for increasingly
unstable cavities the error signal becomes degraded, showing an
offset to the nominal zero crossing, a reduced slope and overall
asymmetry around  the centre. The complete investigation into the
feasibility of such cavities is beyond the scope of this paper, it
includes amongst other issues  the quantitative comparison  of the
control noise from the degradation of the control signals with the
reduced thermal noise. The simulation task described above is
sufficient to provide a test case for our ROQ method.

\section{Application and performance of new integration method} \label{sec:performance}
In this section we provide a detailed and complete recipe for setting up and using
the ROQ for the LIGO example, using \Finesse and \textsc{Pykat}, and
discuss the performance, in terms of speed and accuracy, of our
method. The description should be sufficient for the reader to
implement our method for their own optical setup. In this section we 
include the costly ``offline'' procedure of building the empirical interpolant for completeness.
As part of the ongoing code development of \Finesse and \textsc{Pykat} we intend to pre-generate
the empirical interpolants, suitable for a wide range of problems, so that the typical user should not need
to run the costly offline building of the interpolant into their simulations.

\subsection{Computing the ITM and ETM Empirical Interpolants}
\label{sec:example_compute_EI}
Firstly the range of beam parameters for the simulation must be
determined. Once this is known a training set can be constructed and the empirical interpolant can be computed.
The surface distortions that are of interest are those on the HR surfaces of
a LIGO arm cavity mirror. We will require two EIs, one for the ITM HR
surface and one for the ETM HR surface due to the differing beam parameters at each mirror.
The beam parameter range that the training sets should span
are determined by varying the radius of curvature of the ITM and ETM to include the
range of cavity stabilities which we want to model. The beam parameter ranges are shown in
figure~\ref{fig:param_range}. The required ranges 
for the ETM are $4.7\mathrm{ m} < w_0 < 12.0\mathrm{mm}$ and
$2.11\mathrm{km} < z < 2.20\mathrm{km}$ and for the ITM
$4.7\mathrm{mm} < w_0 < 12.0\mathrm{m }$ and $-1.88\mathrm{km} < z <
-1.79\mathrm{km}$, up to a maximum optical mode order of $\mathcal{O}
= 20$, Netwon-Cotes degree of 6, $L = 1199$. For this example we fix the maximum tolerable error of the empirical interpolant to $\epsilon = 10^{-14}$.

\begin{figure}
      \includegraphics[width=0.48\textwidth]{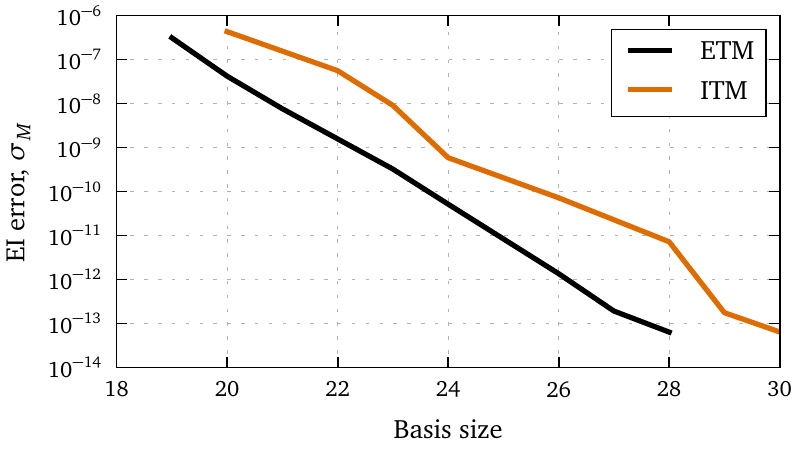}
        \caption{EI error as a function of the number of basis
          elements selected by the greedy algorithm
          (Alg.~\ref{alg:Greedy}) for the example described in
          section~\ref{sec:example_compute_EI}. As expected from the
          error analysis in section~\ref{sec:errors}, the empirical interpolant error
          displays exponential convergence with the basis size.}
        \label{fig:greedy_error}
\end{figure}

Using these ranges the method described in section~\ref{sec:eim} can
be used to produce the EIs. The offline computation of the basis can
have significant computational cost. For very wide parameter ranges
the memory required to store the training sets can quickly exceed that of typical
machines. For the above parameters, with 100 sample points each in the
$w_0$ and $z$ range, up to $\mathcal{O}=14$ and $\epsilon=10^{-14}$
approximately 7GB of memory was required. Running this method
on machines with less memory is possible by storing the training set on
a hard drive using a suitable data storage format such as HDF5 for access.
Computation time of the empirical interpolant is then limited by the read and
write times of the media. Using a MacBook pro 2012 model which
contains a 2.7 GHz Intel core i7 with 8GB of RAM generating the ITM
and ETM reduced basis and empirical interpolant takes $\approx 4$ hours each. The number of elements
in the final reduced basis for the ITM and ETM were $N=30$ and $N=29$
respectively. In figure~\ref{fig:greedy_error} the
convergence of the empirical interpolant error with respect to the acceptable empirical interpolant error. One can see that the
EI error converges exponentially as described in
section~\ref{sec:errors}.

\subsection{Producing the ROQ weights}

\begin{figure*}[htb]
      \includegraphics[width=1\textwidth]{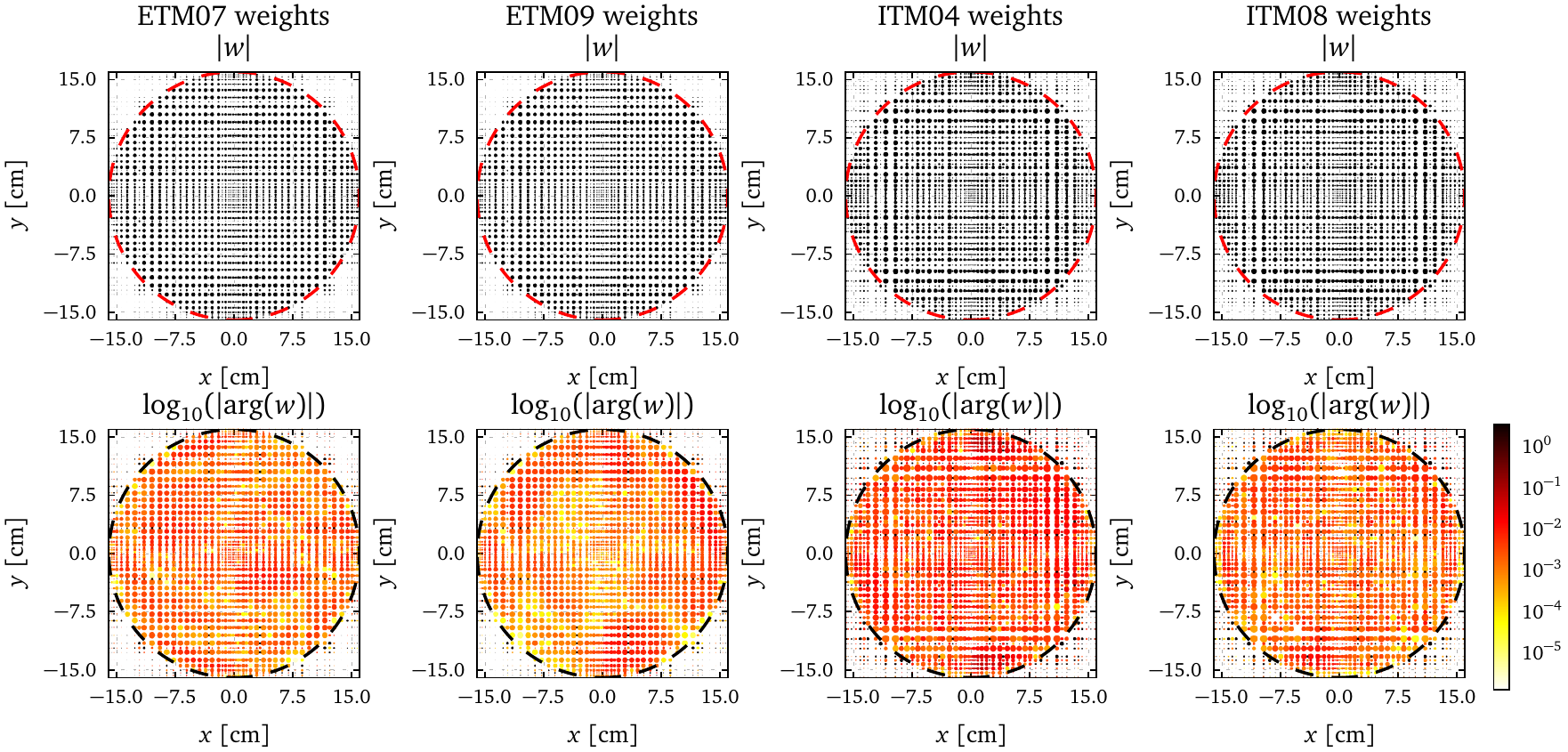}
        \caption{Absolute and argument values of the ROQ weights \eqref{eq:wk} generated
		for each of the maps as shown in figure~\ref{fig:maps}.
		Here the final quadrature rule can be visualized. The top
		plots show the absolute value: The size of the point is proportional to $|w|$ and the center of each point lies on a specific empirical interpolation node in the x-y plane $(X_i, Y_j)$ (c.f. \eqref{eq:interp_problem} and \eqref{eq:roq}).
		The bottom plots show $\log_{10}(arg(w))$. 
		The dashed line on each plot shows the mirror surface boundary; outside the boundary	 the mirror maps are 				equal to zero.
		We note that there are non-zero ROQ weights
		associated with points in the region where the mirror maps are zero. While this may  be counter intuitive, 
		it is a consequence of the fact that the empirical interpolant nodes lie within the full x-y plane and, that they are 				constructed without any knowledge of the mirror maps: the weights still receive no contribution from 						the region where $A(x,y)=0$ as
		this region does not contribute to the sum in \eqref{eq:wk}. However, the ROQ uses information about the kernels \eqref{eq:kernels}
		over the entire region, including where $A(x,y)=0$.
			}
        \label{fig:romified_maps}
\end{figure*}

Once the empirical interpolant has been computed for both ITM and ETM HR surfaces
the ROQ weights~\eqref{eq:wk} can be computed by convoluting
the mirror maps with the interpolant. The surface maps that we have
chosen  are the
measured surface distortions of the (uncoated) test masses currently installed at
the LIGO Livingston observatory, shown in figure~\ref{fig:maps}. The
maps contain $L \approx 1200$ samples 
and we can expect a theoretical speed-up of $L^2/N^2 \approx
1200^2/30^2 = 1600$ from using ROQ over Newton-Cotes. These maps 
include an aperture, $\mathcal{A}$, and the variation in surface
height in meters, $z(x,y)$. Thus to calculate the HOM scattering on
reflection from one of these mirrors with \eqref{eq:knm} the distortion term is: 
\begin{equation}
	A(x,y) = \mathcal{A}(x,y) e^{2\I k z(x,y)} 
	\label{eq:map_distortion}
\end{equation}
where $\mathcal{A}(x,y)$ is 1 if $\sqrt{x^2+y^2} < 0.16$m and 0
otherwise, and $k$ is the wavenumber of the incident optical field.

Using \eqref{eq:map_distortion} with equation~\eqref{eq:wk} (with a Newton-Cotes rule of
the same degree the empirical interpolant was generated with) the ROQ weights can be computed for
each map shown in figure~\ref{fig:maps}.
This computational cost is proportional to the number of elements
in the EI, $M$, and the number of samples in the map, $L^2$. For the LIGO maps this
takes $\approx 10$s on our 2012 MacBook Pro.
The resulting ROQ rule for the maps can be visualised as shown in
figure~\ref{fig:romified_maps}:
the amplitude of the ROQ weights map out the
aperture and the phase of the weights varies for different maps
because of the different surface structure.
The computation
of these ROQ weights need only be performed once for each map, unless
the range of beam parameters required for the empirical interpolant are changed. 

\begin{figure*}
    \centering
    \begin{subfigure}[b]{0.32\textwidth}
        \centering
        \includegraphics[width=\textwidth]{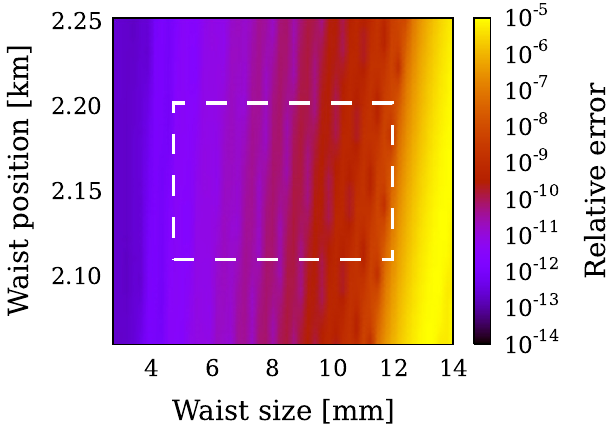}
        \caption{ROQ error for $\mathcal{O}_{max} = 10$}
        \label{fig:param_range_err_10}
    \end{subfigure}
    \hfill
    \begin{subfigure}[b]{0.32\textwidth}
        \centering
        \includegraphics[width=\textwidth]{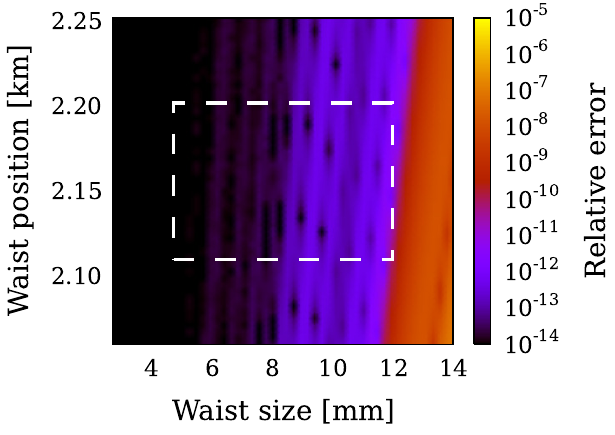}
        \caption{ROQ error for $\mathcal{O}_{max} = 14$}
        \label{fig:param_range_err_14}
    \end{subfigure}
    \hfill
    \begin{subfigure}[b]{0.32\textwidth}
        \centering
        \includegraphics[width=\textwidth]{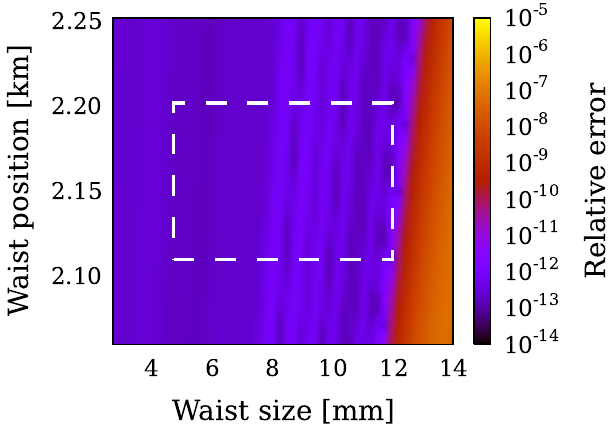}
        \caption{ROQ error for $\mathcal{O}_{max} = 18$}
        \label{fig:param_range_err_18}
    \end{subfigure}
	
    \caption{Maximum relative error between the scattering matrices
      computed for the ETM07 surface map, with ROQ
	and using Newton-Cotes, for mode orders up to  $\mathcal{O}_{max} = 18$. 
	The dashed white area represents the beam-parameter region over which training sets were generated.
	The subplots illustrates how using an ROQ built for a larger
  $\mathcal{O}_{max}$ scattering reduces the maximum error
  significantly. Also shown is that the ROQ is
  valid over a larger parameter range than what it was
  initially generated for, implying that the empirical interpolant can be used for
  extrapolation in a limited parameter region outside the initial range
  indicated by the white dashed box.}	
    \label{fig:param_range_err}
\end{figure*}

\begin{figure}
      \includegraphics[width=0.48\textwidth]{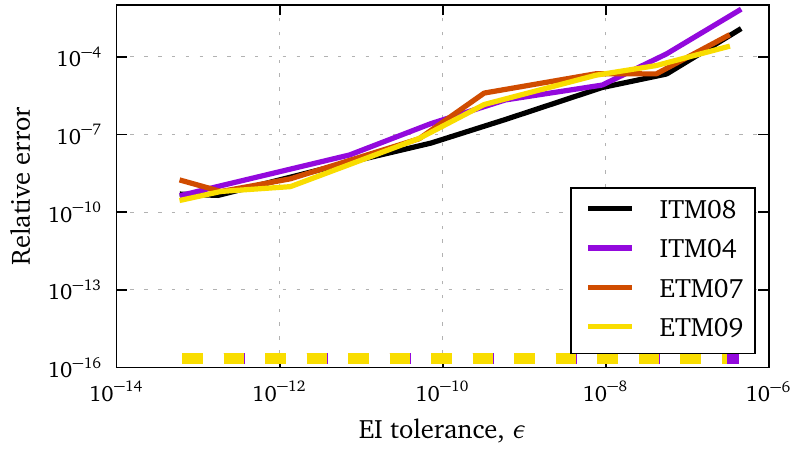}
        \caption{Relative error in the scattering matrices computed using the ROQ
		and Newton-Cotes integration (with $\mathcal{O}_{max} = 10$) as a function of the empirical interpolant tolerance $\epsilon$.
		The empirical interpolant was built for maximum coupling $\maxtem=14$.
		The error is the minimum (dashed lines) and maximum (solid lines)
        over the parameter space with which the empirical interpolant was built for, thus
		represents the worst and best case scenarios. 
		The largest errors are independent of the map data and occur on
		couplings coefficients which couple the higher order modes included in the empirical interpolant.
		}
        \label{tolerance_10_14}
\end{figure}

We verify that the process of generating the ROQ rule has
worked correctly by computing the scattering matrices with ROQ and
Newton-Cotes across the parameter space.
We compute $\hat{k}(q; \mathrm{ETM07})$ with
$\maxtem = 10$ using ROQ and then again using Newton-Cotes
integration. Computing the relative error between each element of these two
matrices the maximum error can be taken for $q$ values spanning the
requested $q$ parameter range. 
Figure~\ref{tolerance_10_14} shows how the final error of the EI,
$\sigma_M$, propagates into  an error in the scattering matrix. This
shows the maximum (solid line) and minimum (dashed line) errors for
any element in the scattering matrix between the two methods. From
this it can be seen that building a more accurate empirical interpolant results in
smaller maximum errors in the scattering matrix. Now, using the most accurate reduced basis
the maximum relative error is shown in
figure~\ref{fig:param_range_err} over the $q$ space, where the white
dashed box shows the boundaries of the parameters in the training set. Overall the method successfully computed a ROQ rule that
accurately reproduced the Newton-Cotes results for scattering up to
$\mathcal{O} = 10$. In should be notes that the largest errors, e.g. as seen in
figure~\ref{tolerance_10_14}, do not represent the full parameter
space but occur only at smallest $z$ and largest
$w_0$. It was also found that
building a basis including a higher maximum HOM, for example basis of
order 14 for scattering computations up to order 10,
significantly improved the accuracy of the ROQ.
Using an reduced basis constructed for order 14 rather than order 10 only increased the number  of
elements in the basis by 2,
thus not significantly degrading any speed improvements. It can also
be seen in figure~\ref{fig:param_range_err} that ROQ extrapolates
beyond the originally requested $q$ parameter space and does not
instantly fail for evaluations outside of it. A gradual decrease in
the accuracy can be seen when using larger $w_0$ values.

\subsection{Performance}

\begin{figure}
      \includegraphics[width=0.48\textwidth]{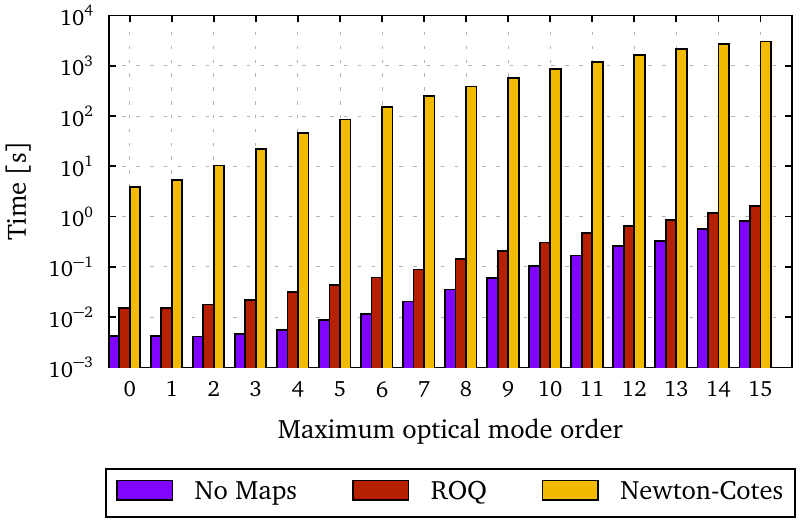}
        \caption{
		Time taken to run \Finesse to model the steady state optical fields
		in an LIGO cavity with surface maps on both the ITM and ETM HR surfaces.
		The timing of running the entirety of \Finesse is used---rather than just the
		core method---because there are additional speed improvements from having to read
		and handle significantly less data points, from the $L\times L$ maps down to $M\times M$ ROQ weights.
		Smaller data fits into processor caches better and also reduces disk read times.
		This plot compares a single computation of the scattering matrices with ROQ \eqref{eq:roq} and Newton-Cotes.
		The case with no maps used is also shown to illustrate how much time
		is spent in \Finesse doing calculations not involving maps, which now becomes
		the dominant computational cost when using ROQ. A significant improvement is also found
		for order zero where only one scattering integral need be calculated;
		this is partly time saved from having to read larger data from the disk and manipulating it
		in memory. The preprocessing is unavoidable as the \Finesse can except many different
		types of map, thus it cannot be optimised at runtime until it know what it is dealing with.
		ROQ helps here as it removes this preprocessing step so it need only happen once.
		The ROQ preprocessing happens during the computing of the ROQ weights~\eqref{eq:wk}.
		This is a one time cost for each map for a particular EI donqe
		outside of \Finesse, thus isn't included in this timing.
		The computational cost of this is on the order of 5s for each map for the reduced basis
		used in this example. 
		} 
        \label{fig:times}
\end{figure}

\begin{figure}
      \includegraphics[width=0.48\textwidth]{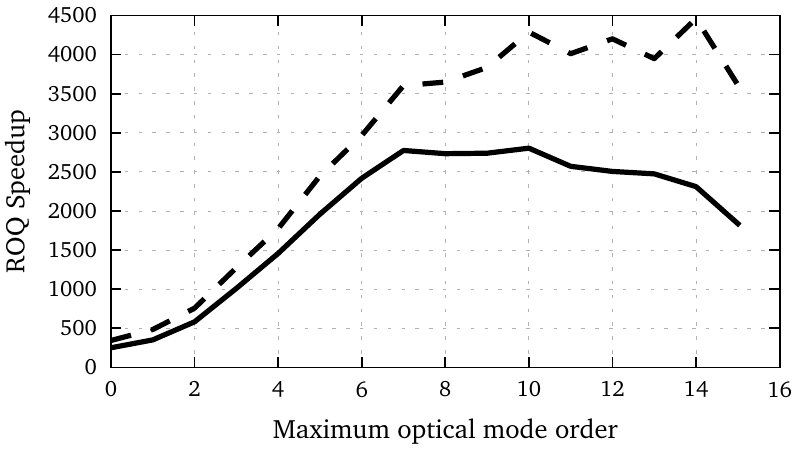}
        \caption{
		The speed-up achieved using ROQ compared to Netwon-Cotes as a function of mode order 
		using the timing values in figure~\ref{fig:times}. The dashed line shows the speed-up
		if the time for initialisation and post-processing
		is subtracted from both times for Newton-Cotes and ROQ. This demonstrates the
		improvements just for the computational cost relating to map scattering calculations.
		Simulations that have a larger computational cost relating to features not related
		to scattering will show a smaller speed-up. For example, the simulations
		results shown in figure~\ref{fig:aligo_cav_instable}
		have a total simulation speed-up of $\approx 80$ but the scattering calculation
		was reduced from $\approx 20.5\,\text{Hours} \rightarrow 30\,\text{s}$.
		}
        \label{fig:speedup}
\end{figure}

The time taken to run
these \Finesse simulations as $\mathcal{O}$ is increased is shown in
figure~\ref{fig:times} demonstrating how much more
efficient it is to use ROQ over Newton-Cotes for the computation of
scattering matrices. We also show for reference the computation time when no scattering from surface maps is included to
give the base time it takes to run the rest of the \Finesse simulation. The overall speed-up achieved can be seen 
in figure~\ref{fig:speedup}, reaching $\approx 2700$ times faster to run the entire simulation at $\mathcal{O}=10$.
The overall speed-up then begins to drop slightly as the base time taken to run the rest of \Finesse becomes larger. The dashed
line in figure~\ref{fig:speedup} shows the speed-up if this base time is removed, again showing an impressive
speed-up peaking at 4000 times faster.

Using ROQ enables us to perform such modelling tasks with a far greater efficiency.
Running the model to compute the output seen in figure~\ref{fig:aligo_cav_instable_scan} 
required computing 100 different scattering matrices for the various
changes in RoC. 
This took 20.5 hours to compute with Newton-Cotes and 18 minutes with ROQ
\footnote{Note that the effective speed-up in this case is less than the large values in
		  figure~\ref{fig:speedup} because here we have included the total runtime of the
		  simulation. This includes the initialisation and running of the other aspects of
		  \Finesse which took $\approx 17$\,minutes. The actual time taken for just the
		  ROQ calculation is $\approx 30\,$s thus a speed-up in the ROQ vs Newton-Cotes
		  is $\approx 2500$.}.
The difference in the final result between ROQ and Newton-Cotes is shown in terms of 
relative error in figure~\ref{fig:aligo_cav_instable_scan_err}. We
have prepared the ROQ input for this example such that the error is significantly
lower than 1\,ppm (relative error of $10^{-6}$) thereby showing that
ROQ can be both much faster and still sufficiently accurate.

\section{Conclusion}

Numerical modelling of optical systems plays a vital role for the
design and commissioning of precision interferometers. The typical use
of the simulation software in this area requires rapid iterations of
many simulation runs and manual fine tuning as modelling progresses,
which is not well suited for large computer clusters.
The scope of current investigations is often limited by
the required computation time and thus the development of fast and flexible tools
is a priority. Current problems in precision interferometers, 
such as LIGO, involve the investigation of laser beam shape
distortions and their effect on the interferometer signal.
Frequency-domain simulations using Gaussian modes to describe
the beam properties have emerged as fast and flexible tools.
However, the computation of the scattering matrix for mirror
surface distortions---effectively an overlap integral of measured surface data
with Hermite-Gauss modes---has shown to be a limiting factor
in improving the computational speed of such tools.
A significant reduction in computational time of current numerical
tools is required for more efficient in-depth modelling of interferometers
including more realistic features such as clipping, optical
defects, thermal distortions and parametric instabilities.

In this work we have demonstrated how the empirical interpolation
method can be used to generate an optimised quadrature rule for
paraxial optical scattering calculations, known as a \textit{reduced order quadrature}. 
Our method removes the
prohibitive computational cost of computing the scattering by speeding
up the calculation of the steady state optical fields in a LIGO arm
cavity by up to a factor of $~2750$ times, reducing simulation
times from days to minutes. Using an exemplary simulation task of
near-unstable arm cavities for the LIGO interferometers  we have
demonstrated that our method is both
accurate and fast for a typical modelling scenario where imperfections
in the interferometer have a significant impact on optical
performance. We have provided a complete recipe to recreate and
use the new algorithm and provide an open source implementation in
our general-purpose simulation software \Finesse.
Importantly, the reduced order quadrature integration method is generic and can be applied
to any optical scattering problem for any surface distortion data.

\section{Acknowledgements}
We would like to thank GariLynn Billingsley for providing the Advanced
LIGO mirror surface maps and for advice and support on using them.
We would also like to thank Peter Deiner, Scott Field and Chad Galley for 
useful discussions and encouragement throughout this project.
Some of the computations were carried out using the high performance computing
resources provided by Louisiana State University (http://www.hpc.lsu.edu) and the Extreme Science and Engineering Discovery Environment (XSEDE) \cite{10.1109/MCSE.2014.80}, which is supported by National Science Foundation grant number ACI-1053575.
This work has been supported by the Science and Technology Facilities
Council and the U.S. National Science Foundation under cooperative agreement NSF-PHY-0757058.
This document has been assigned the LIGO Laboratory document
number LIGO-P1500128-v1.

\bibliographystyle{myplain}
\bibliography{romhom}

\end{document}